\documentclass{article}
\usepackage{spconf}  

\ifdefined\usePortuguese
\usepackage[brazil]{babel}   
\usepackage[latin1]{inputenc}
\else
\fi

\ifdefined\appendices
\else
\ifdefined\addAppendixAsPreamble 
\usepackage[titletoc,title]{appendix} 
\else
\usepackage{appendix} 
\fi
\fi

\usepackage{booktabs}
\usepackage{multirow}
\usepackage{makecell}

\usepackage{makeidx}  
\usepackage{graphicx}
\usepackage{color} 
\definecolor{darkgreen}{rgb}{0,0.35,0}

\ifdefined\keywords
\else

	\def\keywords{\vspace{.5em}{\bfseries\textit{Index Terms}---\,\relax%
	}}
	
\fi

\usepackage{url} 

\ifdefined\RCSfile 
\else
\usepackage{cite} 
\fi

\usepackage{multirow} 

\usepackage{mathbbol}

\ifdefined\amsmath
\else
\usepackage{amsmath}
\fi

\ifdefined\amsfonts
\else
\usepackage{amsfonts}
\fi

\usepackage{amssymb}

\ifdefined\proof
\else
\usepackage{amsthm}
\fi

\ifdefined
\else
\ifdefined\akbook
\ifdefined\lulu
\typearea[6mm]{1} 
\else
\usepackage[a4paper, top=3cm, bottom=3.4cm, left=1.8cm, right=2.5cm]{geometry}
\fi
\fi
\fi

\usepackage{mathrsfs} 
\usepackage{makeidx}  
\usepackage{graphicx,color}

\usepackage{multirow} 
\usepackage{listings} 
\usepackage{ifpdf} 

\ifpdf

\ifdefined\hyperref
\else
\usepackage[pdftex,draft]{hyperref}
\fi

\usepackage{epstopdf}
\hypersetup{%
pdftitle={Some title},
pdfauthor={Aldebaro - LASSE - UFPA},
pdfkeywords={DSP,Signal},
pdfstartview={FitH}, 
urlcolor=black,
linkcolor=black,
citecolor=black,
}

\fi 

\usepackage[english]{babel}
\usepackage[utf8]{inputenc}
\newcommand{\figl}[1] {{Fig.~\ref{fig:#1}}}
\newcommand{\tabl}[1] {{Table~\ref{tab:#1}}}
\newcommand{\email}[1]{\\\mbox{}\\{#1}}
\newcommand{\changed}[1]{{#1}} 

\title{Generating MIMO Channels for 6G Virtual Worlds\\Using Ray-tracing Simulations}

\name{Aldebaro Klautau, Ailton de Oliveira, Isabela Pamplona Trindade and Wesin Alves}
\address{Universidade Federal do Pará - LASSE - www.lasse.ufpa.br\\
		Av. Perimetral S/N, Belém, Pará, Brazil		
		\email{aldebaro@ufpa.br, \{ailton.pinto,isabela.trindade\}@itec.ufpa.br, wesinec@gmail.com}}

\begin{document}




\maketitle

\begin{abstract}
Some 6G use cases include augmented reality and high-fidelity holograms, with this information flowing through the network. Hence, it is expected that 6G systems can feed machine learning algorithms with such context information to optimize communication performance. This paper focuses on the simulation of 6G MIMO systems that rely on a 3-D representation of the environment as captured by cameras and eventually other sensors. We present new and improved Raymobtime datasets, which consist of paired MIMO channels and multimodal data. We also discuss tradeoffs between speed and accuracy when generating channels via ray-tracing. We finally provide results of beam selection and channel estimation to assess the impact of the improvements in the ray-tracing simulation methodology.
\end{abstract}

\section{Introduction}

The 6G systems are expected to support applications such as augmented reality, multisensory communications and high-fidelity holograms~\cite{2021_lima_6g}. 
This information will flow through the network and it is expected that 6G systems will
use machine learning (ML) and, more generally, artificial intelligence (AI), to leverage 
multimodal data and context awareness to optimize performance~\cite{lim_map-based_mmWave_2020}. 
This requires a simulation environment that is capable not only of generating communication channels, but also the corresponding sensor data, matched to the scene. Such simulation that integrates communication networks and artificial intelligence immersed in virtual or augmented reality can be computationally expensive, especially for time-varying digital worlds.

This paper focuses on the simulation
of 6G MIMO systems that rely on a 3-D representation of the environment as captured by
cameras and, eventually, additional modalities of data to make communications more efficient. 
This requirement precludes the adoption of a class of modern channel models that are not related to any virtual world, such as the ones presented in~\cite{wu_general_2018,bian_general_2021}.
The motivations for using ray-tracing (RT) in this work are aligned with the trends detailed in~\cite{lim_map-based_mmWave_2020}.
Also, the paper discusses tradeoffs between speed and
accuracy when generating channels via RT.
Controlling these tradeoffs by tuning the simulations is important to 
facilitate research in practical ML-based optimizations of the 6G physical layer.
Investigations in this area still need to depart from using relatively small datasets,
to benefit from deep learning and other techniques that perform best in the large data regime. 

Due to characteristics such as large antenna arrays in 6G ultra-massive MIMO systems and higher frequency bands, 6G measurement campaigns will require expensive equipment in order to be performed. Proper simulation methodologies for generating communication channels are important to generate abundant data in controlled conditions and foster the adoption of ML/AI for optimizing the 6G physical layer.
Given the availability of a virtual world representation in some 6G use cases, this paper 
concerns frameworks based on RT that support
\emph{Communication networks and Artificial intelligence immersed in VIrtual or Augmented Reality} (CAVIAR). 
The contributions of this paper are:
\begin{itemize}
	\item Two new and improved Raymobtime datasets~\cite{klautau_5g_2018} that maintain aligned the orientations of the MIMO antenna array and the vehicle in which the array is installed.
	\item Support to automatically generate MIMO channels without resorting to the geometric channel model, which assumes planar wave propagation and may be inadequate to 6G large antenna arrays.
		\item Source code and datasets to reproduce the results of this paper.\footnote{\url{https://github.com/lasseufpa/SSP-Raymobtime}.}
\end{itemize}

The paper is organized as follows: Section~\ref{sec:caviar} describes 
characteristics and requirements of CAVIAR frameworks.
Section~\ref{sec:ray_tracing} explains the improvements in the RT simulation methodology.
Section~\ref{sec:results} presents numerical results and their discussion. Finally, Section~\ref{sec:conclusions} concludes the paper.

\section{CAVIAR Simulation Requirements}\label{sec:caviar}

\begin{table*}[hbt]
	\centering
	\caption{Some Raymobtime datasets and the new ones: s011 and s012. 
	\label{tab:datasets}}		
	\resizebox{\textwidth}{!}{%
	\begin{tabular}{cccccccc} 
		\hline
		Dataset name & 
		\begin{tabular}[c]{@{}c@{}}Frequency \\ (GHz) \end{tabular} & 
		\begin{tabular}[c]{@{}c@{}}Number of receivers \\and type \end{tabular} & 
		\begin{tabular}[c]{@{}c@{}}Time between \\scenes (ms) \end{tabular} & 
		\begin{tabular}[c]{@{}c@{}}Time between \\episodes (s) \end{tabular} & 
		\begin{tabular}[c]{@{}c@{}}Number of \\episodes \end{tabular} & 
		\begin{tabular}[c]{@{}c@{}}Number of scenes \\per episode \end{tabular} & 
		\begin{tabular}[c]{@{}c@{}}Number of valid \\channels\end{tabular} \\ 
		\hline
		s001 & 60 & 10 Mobile & 100 & 30 & 116 & 50 & 41 K \\
		s005 & 2.8 and 5 & 10 Fixed & 10 & 35 & 125 & 80 & 100 K \\
		s006 & 28 and 60 & 10 Fixed & 1 & 35 & 200 & 10 & 20 K \\
		s008 & 60 & 10 Mobile & - & 30 & 2086 & 1 & 11 K \\
		s011 (new) & 60 & 10 Mobile & 500 & 6 & 76 & 20 & 13K \\
		s012 (new) & 60 & 10 Fixed & 500 & 6 & 105 & 20 & 21K \\
		\hline
		\end{tabular}
	}
\end{table*}



There are many applications of augmented and virtual reality, for instance, in education and gaming. 
We are interested in a special class of such applications, which incorporates a communication subsystem and also AI/ML. 
A CAVIAR framework is based on the availability of a description $\Gamma_t$ of a digital world at discrete time $t \in \mathbb{Z}$.
In this paper, we focus on a concrete example in which the communication subsystem is a MIMO digital transmission model and AI/ML is used for beam selection and channel estimation. The nomenclature for this MIMO use case is discussed in the next paragraphs.

For a given time instant (\emph{scene}), multicarrier (OFDM, etc.) MIMO systems require a set $\mathcal{H} = \{\textbf H_{k}\}, k=1,\ldots,K$ of \changed{$N_\mathrm{rx} \times N_\mathrm{tx}$} complex-valued matrices, where $k$ is the subcarrier index, and \changed{$N_\mathrm{rx}$ and $N_\mathrm{tx}$} are the number of antenna elements at the receiver and transmitter, respectively. For narrowband systems, $K=1$ and $\mathcal{H} = \{\textbf H\}$.
For time-varying discrete-time channels, $\mathcal{H}_t = \{\textbf H_{t,k}\}$ denotes the channel at time $t$. Given a sampling interval $T$ and an \emph{episode} with $S$ scenes, the sequence $[\mathcal{H}_1, \ldots, \mathcal{H}_S]$ indicates the channel evolution over time, as defined in the \emph{Raymobtime} methodology~\cite{klautau_5g_2018}, and suffices to describe the relations among channel inputs and outputs when conventional signal processing is adopted.

A CAVIAR simulation requires not only the matrices in $\mathcal{H}_t$ but also the corresponding parameters set
$\mathcal{P}_t = \{\Theta_{{t},{p}}\}, p=1,\ldots,P$, with $P$ being the number of distinct sets of parameters (or \emph{features}) extracted from the scene description $\Gamma_t$ from one or more sensors.
 For instance, $P=1$ was adopted in \cite{klautau_LIDAR_paper}, which assumed the raw data from a LIDAR sensor was converted to features $\Theta_{{t},{1}}$. In \cite{our_paper_with_multimodality}, $P=3$ distinct feature modalities were adopted: $\Theta_{{t},{1}}$ with positions from a GNSS (GPS) receiver, $\Theta_{{t},{2}}$ with resampled images from RGB cameras and $\Theta_{{t},{3}}$ with features from LIDAR~\cite{URL-COM-FIGURA-DO-ITU-Challenge}. 

\begin{figure}[htb]
\includegraphics[width=\columnwidth]{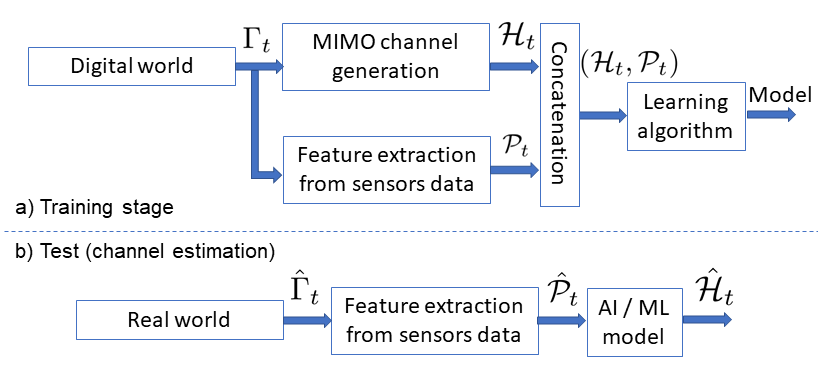}%
\caption{Example of CAVIAR framework for applying AI/ML to 6G MIMO systems based on multimodal data during: a) the training and b) test stages. For concreteness channel estimation is assumed and during the test stage, the AI/ML model generates a channel estimate $\hat{\mathcal{H}}_t$ based solely on features $\hat{\mathcal{P}}_t$ extracted from real-world sensors.
	\label{fig:caviar}}
\end{figure}

As depicted in Fig.~\ref{fig:caviar}, the \emph{paired} information $(\mathcal{H}_t, \mathcal{P}_t)$ enables the generation of AI/ML models based on supervised learning applied to tasks such as power allocation or channel estimation.
Assuming the latter, during its test stage (Fig.~\ref{fig:caviar}b), the trained AI/ML model estimates the channel $\hat{\mathcal{H}}_t$ based on features $\hat{\mathcal{P}}_t$ extracted from a real world representation $\hat{\Gamma}_t$.

The 6G AI/ML-based algorithms will leverage additional context information and the CAVIAR framework must support pairing channel datasets with other data modalities.
A suitable dataset for CAVIAR follows the Raymobtime methodology and organizes the data in $E$ episodes, each one with a sequence $[(\mathcal{H}_1, \mathcal{P}_1), \ldots, (\mathcal{H}_S, \mathcal{P}_S)]$ of paired data, in the sense that all information in $(\mathcal{H}_t, \mathcal{P}_t)$ corresponds to the scene at time $t$ in the virtual world.
Each episode can consist of a smooth \emph{trajectory} or \emph{snapshots} spaced apart in time. The first category of dataset is useful for applications such as channel tracking, in which previous AI/ML decisions influence the current one. For trajectories, $T$ is relatively small, the scenes are similar and the channels $\mathcal{H}_t$ present significant correlation. In contrast, datasets with snapshots are suitable to problems such as initial channel acquisition and
are often designed with $T$ large enough to have diverse scenes (and, consequently, a variety of channels). For instance, dataset s005 has 100K MIMO channels of trajectories, \changed{with each episode containing $80$ scenes} spaced by $T=10$~ms. On the other hand, dataset s008 has snapshots with each episode containing just $1$ scene. Table~\ref{tab:datasets} shows some characteristics of some Raymobtime datasets. 


In order to develop CAVIAR frameworks for 6G, it is important to address issues related to the tradeoff between computational cost and realism of RT simulations. The next section presents two improvements toward more realistic datasets for AI/ML involving MIMO channels.

\section{Methodology Improvements}
\label{sec:ray_tracing}


Regarding the MIMO channel generation, the variety of use cases in 5G and 6G promote a large number of deterministic and stochastic channel models, such as: measurement-based, map-based, point-cloud and geometry-based stochastic models~\cite{lim_map-based_mmWave_2020}.
By definition, the CAVIAR framework relies on a description of a digital world $\Gamma_t$, and ray-tracing models fit naturally given that the MIMO channels $\mathcal{H}_t$ are obtained from the corresponding 3-D scene.

As applied to channel modeling, RT aims at predicting the effect of buildings, vehicles, 
topographic features and other obstacles on the propagation of electromagnetic fields. The geometry and materials that constitute a given 3-D environment are defined prior to the simulation. Then, the transmitters and receivers 
are positioned inside in the scene. The algorithm shoots rays that realistically propagate through the geometry 
using the uniform theory of diffraction (UTD) and is able to calculate each ray paths 
electromagnetic fields. The information from all ray paths is then processed to provide parameters such 
as received power, path loss, and MIMO channel matrix. 
In this work, Remcom's Wireless InSite (WI) RT software~\cite{REMCOM_WI_2019} was adopted given its widespread use~\cite{lim_map-based_mmWave_2020}.





Given a RT simulator, among many existing alternatives~\cite{lim_map-based_mmWave_2020}, this paper discusses three procedures for generating MIMO channels that have been used in AI applications. The first one consists in incorporating the antenna array(s) within the RT simulation. The second alternative uses only SISO systems with isotropic antennas, and the \emph{geometric} channel model~\cite{sayeed_deconstructing_2002,tse2005fundamentals} is then used in a post-processing step to convert the obtained RT information into MIMO channels. In~\cite{trindade2020accuracy}, these two alternatives are contrasted for different scenarios, and here they are called MIMO-RT and SISO-RT-GEO, respectively. Assuming a narrowband channel at time $t$, the SISO-RT-GEO channel is obtained after the RT finishes via the geometric channel as follows~\cite{tse2005fundamentals}:

\begin{align}
\textbf H_{t}  = \sqrt{\changed{N_\mathrm{rx} N_\mathrm{tx}}}\sum_{\ell = 1}^L \alpha_{\ell} \mathbf{a}_{\textrm{rx}}(\phi_\ell^A, \theta_\ell^A)\mathbf{a}^H_{\textrm{tx}}(\phi_\ell^D, \theta_\ell^D), 
\label{eq:narrow_geo_model}
\end{align}
where the number $L$ of multipath components (MPCs), 
the gain $\alpha_\ell$ and the angles $\phi$ and $\theta$ of the steering vectors $\mathbf{a}_{\textrm{tx}}$ and $\mathbf{a}_{\textrm{rx}}$, are all determined by RT using isotropic antennas at transmitter and receiver, as explained e.\,g. in~\cite{trindade2020accuracy}.

When using the WI RT simulator, creating a script repeatedly obtain SISO-RT-GEO channels is easier than MIMO-RT. For SISO-RT-GEO, the RT data required by Eq.~(\ref{eq:narrow_geo_model}) is written in ASCII text file (extension p2m) and can be conveniently post-processed. However, when MIMO-RT is adopted in WI, the text files are not generated due to the large amount of data (received power, complex impulse response, angle of arrival and departure, etc.), and 
SQLITE files are used instead. For this paper, we had then to write new Python code to repeatedly run a simulation with antenna arrays, properly retrieve the results \changed{from} the SQLITE file and calculate the MIMO-RT channel matrix via another post-processing step.


The third alternative addressed here is to adopt the geometric channel of Eq.~(\ref{eq:narrow_geo_model}) but obtain the parameters from random distributions instead of RT. We call this procedure RDM-GEO and it is used, e.\,g., in~\cite{jin2020,ma_data-driven_2020,ma_sparse-channel_2020} to assess ML solutions for channel estimation in mmWave MIMO systems.
In particular, $L=3$ and 2 MPCs are used in~\cite{jin2020} and \cite{ma_sparse-channel_2020}, respectively, while six clusters with $L=10$ MPCs each are used in \cite{ma_data-driven_2020}. 

With respect to accuracy, the geometric channel model, and consequently both SISO-RT-GEO and RDM-GEO, assume planar wave propagation~\cite{medbo_radio_propagation_2016}, which may be invalid, for instance in 6G ultra-massive MIMO~\cite{Trends_Wang_2020}. MIMO-RT is not limited by this assumption~\cite{REMCOM_WI_2019}.
Regarding computational complexity, 
SISO-RT-GEO is faster than MIMO-RT and consumes less storage space. RDM-GEO is much simpler than both, given that it does not require a RT simulation.
Therefore, in the next section, we compare the three procedures and their impact in 6G applications that rely on the generated MIMO channels.




\begin{figure}[htb]
\begin{center}
	\includegraphics[width=4.5cm]{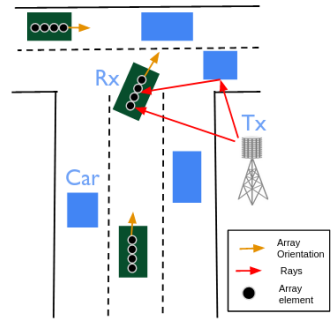}%
	\caption{Illustration of maintaining the correct orientation of the array (ULA with 4 antenna elements in this example) when the orientation of the mobile \changed{receiver} (vehicles, represented by rectangles) is changed.\label{fig:orientation}}
\end{center}
\end{figure}

Another improvement in the RT methodology was to maintain aligned the orientations of the MIMO antenna array and the vehicle in which the array is installed. \figl{orientation} depicts the issue.
As mobile objects (vehicles, people, etc.) move in the virtual world, previous versions of Raymobtime datasets were not updating the orientation of the antenna array. Now, this is done for both SISO-RT-GEO and MIMO-RT, and will be evaluated with ML experiments.

\section{Numerical Results}\label{sec:results}

The first results concern a comparison of the computational cost when executing MIMO-RT and SISO-RT-GEO.
As described in \tabl{cost}, we used datasets s011 and s012 assuming uniform linear array (ULA) with 
64 antenna elements on the base station (BS) and 8 on the user equipment. The simulations were performed on NVIDIAs GeForce RTX $2070$ and RTX $2080$ super, for s012 and s011, respectively. 
The analyzed parameters were the RT simulation time and size of each simulation snapshot. 
The values in \tabl{cost} represent \changed{an average among the simulations for each channel in the dataset}, and the MIMO-RT results are relative to the SISO-RT-GEO results. For instance, when using s011, the RT simulation time was 1.4 longer for MIMO-RT than for SISO-RT-GEO, with the latter one corresponding to a duration of 29.52 seconds in average.

\begin{table}
	\centering
	\caption{Computational cost when executing MIMO-RT and SISO-RT-GEO.\label{tab:cost}}
	\resizebox{\columnwidth}{!}{%
	\begin{tabular}{ccccc}
		\hline
		Type of receiver               &
		 Modeling  & 
		 \thead{Size of\\output files} & 
		 \thead{Simulation\\time} & 
		 \thead{Post-processing\\time} \\ \hline
		\multirow{2}{*}{Fixed (s012)}  & SISO-RT-GEO       & 1 (\changed{1.04 MB})    & 1 (\changed{15.69} s)     & 1 (\changed{0.35} s)           \\
									   & MIMO-RT & \changed{118} x           & 2.2 x           & \changed{7.8} x                \\ \hline
		\multirow{2}{*}{Mobile (s011)} & SISO-RT-GEO       & 1 (\changed{2.23 MB})    & 1 (\changed{29.52} s)     & 1 (\changed{0.71} s)           \\
									   & MIMO-RT & 77 x            & 1.4 x           & \changed{5} x   \\ \hline            
	\end{tabular}%
		}
\end{table}

\begin{figure}[htb]
	\includegraphics[width=7cm]{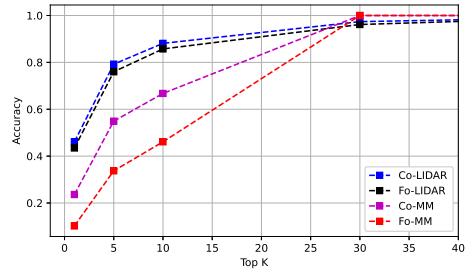}%
	\caption{Top-$K$ accuracy of beam selection for NNs with different inputs --- LIDAR or multimodal (MM) --- and using datasets with corrected orientation (Co) or not (Fo).\label{fig:beam_selection}}
\end{figure}

The results in \figl{beam_selection} are for beam selection, which is already a classical application of ML to the physical layer~\cite{klautau_5g_2018,lim_map-based_mmWave_2020}. In this experiment we followed the procedure proposed in \cite{klautau_LIDAR_paper}, and dense neural networks (NNs) were used to select the best beam pair among 256 possible pairs of indices of a codebook, assuming an analog MIMO architecture. The input to this network are features $\Theta_{{t},{1}}$ extracted from a LIDAR simulator for two plots. The other two plots are obtained with a distinct NN architecture, which uses multimodal (MM) data (LIDAR, position, and images) as inputs~\cite{our_paper_with_multimodality}. Each of these two architectures were used with two distinct versions of the s008 dataset: one with the new and correct orientation (CO), while the other antenna arrays had a fixed orientation (FO) regardless any vehicle rotation.
The top-$K$ accuracy was adopted as figure of merit, and the abscissa indicates the $K$ values. It can be seen that correcting the orientation impacted the NN with a multimodal input but not the one that relies on the LIDAR only. An interesting result is that the multimodal input was beneficial for large values of $K$, and allowed the NN classifier to reach 100\% of accuracy when $K \ge 30$. The details about the NNs and other parameters can be found in the provided source code.

\begin{figure}[htb]
	\includegraphics[width=\columnwidth]{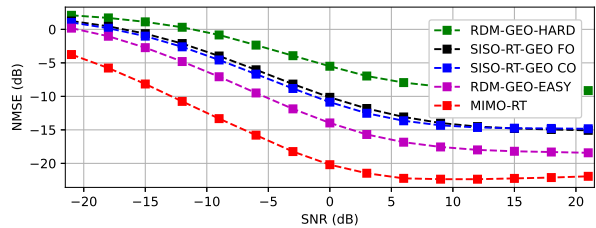}%
	\caption{Channel estimation based on 1-bit MIMO systems.\label{fig:channel_estimation}}
\end{figure}

\figl{channel_estimation} depicts the results of channel estimation using MIMO systems adopting analog to digital converters with resolution of a single bit. 
For this task, we used convolutional NNs as described in~\cite{alves_dtl_mimo} and the companion source code.
ULAs with \changed{$N_{\mathrm{tx}} = 64$ and $N_{\mathrm{rx}} = 8$} antennas were adopted. The normalized mean-squared error 
(NMSE)~\cite{alves_dtl_mimo} is the figure of merit. We first compared SISO-RT-GET with (CO) and without (FO) orientation correction, and both led to similar results. We then trained and tested a NN with data using MIMO-RT, which led to a significantly lower NMSE. 
For instance, the NN trained and tested with the MIMO-RT dataset outperforms the SISO-RT-GEO FO by 10 dB at $\textrm{SNR} = 0$.
Two versions of RDM-GEO were also compared. The first version, RDM-GEO-HARD, used MIMO channels composed by $L=2$ MPCs with gains obtained from Gaussian distributions and all angles from uniform distributions with support $[0, 360[$ degrees. The RDM-GEO-EASY had the angles distributed with an angular spread of 3 degrees around nominal values. \figl{channel_estimation} shows the strong impact of the distributions on the ML result.



The presented results and the ones in literature indicate that the datasets must be carefully chosen when used to assess ML-based approaches.
For example, in~\cite{zhang_deep_2020,dong_cgan_2020} the data from a single indoor scenario is shuffled and split into training and test sets. 
This leads to a situation similar to speaker dependencies in automatic speech recognition (ASR). For instance, as discussed in \cite{zhu_2009_adaboost}, the assessment of a ML algorithm on the \emph{vowel} dataset uses a test set with speakers that are not part of the training set, otherwise overly optimistic results are obtained. 

\section{Conclusions}\label{sec:conclusions}

This paper presented two improvements in generating datasets for 6G MIMO systems that rely on ray-tracing and give support to multimodal paired data. It also introduced the concept of \emph{CAVIAR simulations}.
The results with two distinct applications of ML, beam selection and channel estimation, allowed to observe the impact of the improvements in the ray-tracing simulation methodology, and the importance of proper datasets when evaluating ML-based algorithms to avoid unfair comparisons to conventional signal processing. Besides, aiming at realistic simulations is the natural path to gain better understanding on how ML/AI can make communication systems more efficient.



\bibliographystyle{IEEEtran}
\bibliography{references}


\end{document}